\documentclass[12pt,preprint]{aastex6}
\usepackage{natbib}
\usepackage{graphicx}
\usepackage{amssymb}
\usepackage{amsmath}
\shorttitle{The PBH fraction in cool-core galaxy clusters}
\begin{document}
\title{The evaporating primordial black hole fraction in cool-core galaxy clusters}
\author{Chak Man Lee, Man Ho Chan}
\affil{Department of Science and Environmental Studies, The Education University of Hong Kong, Hong Kong, China}
\email{chanmh@eduhk.hk}

\begin{abstract}
Recent studies of gamma-ray, cosmic-ray and radio data put stringent constraints on the fraction of primordial black holes (PBHs) in our universe.  In this article, we propose a new indirect method in using the X-ray luminosity data of cool-core clusters to constrain the evaporating PBH fraction for the monochromatic, log-normal and power-law mass distributions. The present results show that the amount of evaporating PBHs only constitutes a minor component of dark matter for a large parameter space. The constraints are consistent with and close to that obtained from other cosmic-ray and multi-wavelength observations.
\end{abstract}


\section{Introduction}
Observational data indicate that there is some missing mass in our universe. Some new theoretical particles have been proposed to account for the missing mass, such as weakly interacting massive particles (WIMPs). However, no such particles have been discovered so far in both direct detections \citep{Aprile,Abecrcrombie,Chan9} and indirect detections \citep{Ackermann,Albert,Chan3,Egorov,Chan4,Chan5,Chan6,Chan7,Boudaud2,Bergstrom,Cavasonza}. On the other hand, previously some studies have suggested that overdensities in early universe would collapse to form primordial black holes (PBHs) \citep{Zeldovich,Hawking2}. These PBHs might be able to account for at least some of the missing mass in our universe.

Recent merger events detected from the Laser Interferometer Gravitational Wave Observatory (LIGO) may indicate that some massive PBHs ($\sim 1-100M_{\odot}$) exist \citep{Bird}. Nevertheless, it is also possible for PBHs to have mass less than $1M_{\odot}$. Theoretically, PBHs would emit Hawking radiation and give out high-energy particles such as photons, electrons and positrons \citep{Hawking,Boudaud}. This effect would be much more significant for smaller PBH mass. Therefore, detection of these particles could be an effective way to verify the PBH scenario, especially for the evaporating PBH mass range $\sim 10^{14}-10^{17}$ g. In particular, various kinds of data such as gamma-ray \citep{Carr2,Laha}, cosmic-ray \citep{Boudaud}, radio \citep{Chan8} and strong lensing \citep{Montero} data have been considered to constrain the amount of the evaporating PBHs. The fraction of PBHs being dark matter $f$ is much less than 1 for a large window of the evaporating PBH mass \citep{Smyth}. For the non-evaporating PBH mass range ($>10^{17}$ g), other techniques such as gravitational lensing or dynamical methods can be used to constrain the PBH fraction \citep{Carr4}.

In view of the constraints, less attention has been paid in the X-ray bands \citep{Carr3}. It is because the X-ray contribution of PBHs is difficult to be differentiated from the diffuse background so that the constraints would be much less stringent. Some previous studies have suggested to use the spatial coherence of the X-ray and infrared source-subtracted background to constrain the PBH fraction \citep{Cappelluti,Carr3}. Besides, the future accurate hard X-ray detection might also be able to fulfill the task and give good constraints for PBHs \citep{Ballesteros}. Fortunately, before getting an accurate X-ray detection, we could constrain PBH fraction indirectly by existing X-ray data.  In this article, we discuss a new independent method to constrain the PBH fraction $f$, mainly in the evaporating PBH mass range. By using the X-ray data of some cool-core galaxy clusters, we can constrain the PBH fraction $f$ in these galaxy clusters, and we find that the results are comparable to the gamma-ray, cosmic-ray and radio constraints. It could provide a complementary analysis in constraining the PBH fraction for the evaporating mass range.

\section{Dark Matter Annihilation for the PBH evaporation model}
Similar to traditional black holes, PBHs with mass $M_{\rm PBH}$ emit Hawking radiation and particles. The primary emission spectrum of electron and positron pairs can be given by the following distribution \citep{Boudaud}
\begin{equation}
\frac{d^2N_e}{dEdt}=\frac{\Gamma_e}{2\pi}\left[\exp\left(E/k_{\rm B}T\right)+1\right]^{-1},
\end{equation}
where the temperature $T$ as a unique parameter linking with their mass $M_{\rm PBH}$ dictates the rate of the particles evaporated from the PBHs' surface and is given by \citep{Hawking,Hawking1}
\begin{equation}
k_{\rm B}T=\frac{\hbar c^3}{8\pi G M_{\rm PBH}}\approx 1.06\left(\frac{10^{13}{\rm g}}{M_{\rm PBH}}\right) {\rm GeV}.
\end{equation}
The electron absorption probability $\Gamma_e$ can be approximately modelled as \citep{MacGibbon}
\begin{equation}
\Gamma_e\approx\left\{
\begin{array}{ll}
16 \frac{G^2M_{\rm PBH}^2E^2}{\hbar^3c^6}\,\,\,\,{\rm for}\,\,\,\, \frac{GM_{\rm PBH}E}{\hbar c^3}< 1,\\
27 \frac{G^2M_{\rm PBH}^2E^2}{\hbar^3c^6}\,\,\,\,{\rm for}\,\,\,\, \frac{GM_{\rm PBH}E}{\hbar c^3}\ge 1.
\end{array}\right.
\end{equation}
Apart from the primary emission, PBH evaporation would also give quark pairs which would quickly hadronize to produce secondary electrons and positrons. The spectrum of the secondary electron-positron pairs produced can be calculated by the BlackHawk code published in \citet{Arbey}. We will consider the total emission as the sum of the primary and secondary emissions.

In the followings, we will consider two classes of mass distribution of PBH: monochromatic mass distribution and extended mass distribution. Starting from the former case, we assume that all PBHs have a common mass (i.e. the monochromatic mass distribution). The number of injected electrons per unit time, energy and volume is given by
\begin{equation}
Q(E,r)=\frac{\rho_{\rm PBH}(r)}{M_{\rm PBH}}\frac{d^2N_e}{dEdt}.
\label{monochromatic}
\end{equation}
where $\rho_{\rm PBH}(r)$ denotes the total PBH mass density, which is a function of the distance from the cluster centre, $r$. Assume that the total PBH mass density traces dark matter density $\rho_{\rm DM}(r)$ \citep{Carr}, the total PBH density $\rho_{\rm PBH}(r)$ can be represented by $f\rho_{\rm DM}(r)$, with the PBH fraction $f$.

Beside the monochromatic mass distribution, we generalize the case of extended mass distribution for a more realistic situation. The number of injected electrons per unit time, energy and volume can be written as
\begin{equation}
Q(E,r)=\frac{ \rho_{\rm PBH}(r)}{\rho_{\odot}}\int_{\forall M_{\rm PBH} } dM_{\rm PBH}\frac{g(M_{\rm PBH})}{M_{\rm PBH}}\frac{d^2N_e}{dEdt},
\label{extended}
\end{equation}
where $g(M_{\rm PBH})$ is any particular mass distribution of PBH normalized to $\rho_{\odot}$. $\int_{\forall M_{\rm PBH} } dM_{\rm PBH}$ indicates the integration over the entire possible PBH mass range during formation. Specifically, we focus on two different popular extended mass distributions: log-normal distribution and power-law distribution.

The log-normal distribution \citep{Boudaud,Green,Laha2,Bellomo,Alexandre}, with $\mu$ and $\sigma$ denoted as the median and the standard derivation of the logarithm of the mass distribution respectively, is defined as \citep{Krishnamoorthy}
\begin{equation}
g(M_{\rm PBH})=\frac{\rho_{\odot}}{\sqrt{2\pi}\sigma M_{\rm PBH}}\exp\left[-\frac{\ln^2(M_{\rm PBH}/\mu)}{2\sigma^2}\right],
\end{equation}
whereas the power law distribution, as parameterized by the power law index, $p$, the maximum value $M_{\rm max}$, and the minimum value $M_{\rm min}$ of the mass distribution, is defined as \citep{Laha2,Bellomo}
\begin{equation}
g(M_{\rm PBH})=\frac{p \rho_{\odot}}{M_{\rm max}^{p}-M_{\rm min}^{p}}M_{\rm PBH}^{p-1},
\end{equation}
where $M_{\rm PBH}\in [M_{\rm min}, M_{\rm max}]$ and the power index $p	\neq 0$.

If there exists a large amount of PBHs in a galaxy cluster, the evaporation of PBHs would give a large amount of high-energy electrons and positrons ($\sim 0.1-100$ MeV) with Lorentz factor $\gamma$. When high-energy electrons and positrons are produced, they would diffuse and cool down via the diffusion-loss equation \citep{Atoyan,Chan1}.
The cooling of electrons and positrons is dominated by four processes: synchrotron, redshift-dependent inverse Compton scattering (ICS), bremsstrahlung and Coulomb losses. The total cooling function $b_T(E,r)$ for such four processes can be explicitly expressed (in unit of $10^{-16}$ GeV s$^{-1}$) by \citep{Colafrancesco}
\begin{eqnarray}
b_T(E,r)&=&0.0254E^2B^2 \nonumber\\
&&+ 0.25E^2(1+z)^4 \nonumber\\
&&+ 1.51 n(r)\left[0.36+\log\left(\frac{\gamma}{n(r)}\right)\right]\nonumber\\
&&+6.13 n(r)\left[1+\frac{1}{75}\log\left(\frac{\gamma}{n(r)}\right)\right],
\label{cooling}
\end{eqnarray}
where $z$, $B$ and $n(r)$ are the redshift, magnetic field strength and thermal electron number density of the galaxy cluster, respectively. Here, $n(r)$, $E$, $B$ are in the units of ${\rm cm}^{-3}$, GeV and $\mu$G, respectively. The number density of the hot gas particles (thermal electrons) $n(r)$ is usually constructed by the single-$\beta$ profile \citep{Chen}
\begin{equation}
n(r)=n_0\left(1+\frac{r^2}{r_c^2}\right)^{-3\beta/2},
\label{nodensity}
\end{equation}
where $n_0$, $r_c$ and $\beta$ are empirical parameters fitted from the surface brightness profile.

For most of the galaxy clusters, the cooling timescale is much shorter than the diffusion timescale \citep{Colafrancesco}. Therefore, the equilibrium state solution for the diffusion-loss equation is then expressed, in terms of $Q(E,r)$ [Eqs.~(\ref{monochromatic}) and (\ref{extended})], by \citep{Colafrancesco,Storm}
\begin{equation}
\frac{dn_e}{dE}=\frac{1}{b_T(E,r)}\int_E^{\infty} Q(E',r)dE',
\end{equation}
where $n_e$ is the number density of the high-energy electrons or positrons emitted by PBH evaporation.

In fact, in galaxy clusters, the cooling of electron and positron pairs via synchrotron emission would emit electromagnetic waves in radio bands. These radio waves would finally leave the galaxy clusters. For ICS and Bremsstrahlung cooling, the emission would be mainly in hard X-ray or gamma-ray bands, which would also leave the galaxy clusters. Nevertheless, for Coulomb loss cooling, the energy of the electron and positron pairs would transfer to the thermal electrons and protons in the hot gas. In other words, the evaporation of PBHs would indirectly transfer the energy to the hot gas via Coulomb loss only. Therefore, the total power gain of the hot gas within a radius $R$ in the core of a galaxy cluster is given by
\begin{equation}
L=2\int_0^R\int_{m_e}^{\infty}b_C(E,r)\frac{dn_e}{dE}dE(4\pi r^2)dr,
\label{luminosity}
\end{equation}
where $b_C(E)$ is the power gain via the Coulomb interactions between the high-energy electron-positron pairs and the thermal electrons and protons in the hot gas (the fourth term in Eq.~(\ref{cooling})). The factor of 2 in the above equation indicates the contributions of both high-energy electrons and positrons.

\section{Results}
We have chosen four cool-core galaxy clusters (A262, A2199, A85 and NGC5044) for analysis. Using cool-core clusters is a good option because their relatively low temperature at the central core regions can give stringent constraints for PBH evaporation rate. The four chosen cool-core clusters have low core temperature ($T_c<3$ keV) and very small sizes of cores ($<3$ kpc) \citep{Sanderson}, which suggest that they are very good targets for analysis.

For galaxy clusters, the density of dark matter is usually modelled by the Navarro-Frenk-White (NFW) profile
\begin{equation}
\rho_{NFW}=\rho_s\frac{r_s}{r\left(1+\frac{r}{r_s}\right)^2}
\end{equation}
via its two scale parameters: the scale radius $r_s$ and scale density $\rho_s$ \citep{Navarro}. The scale parameters of the four cool-core galaxy clusters can be obtained in \citet{Chan1}. For the small cool-core regions, we have $r \ll r_s$ so that $\rho_{NFW} \approx \rho_sr_s/r$. Note that the NFW profile may not be very good at representing the dark matter density near the central region of a galaxy cluster \citep{Sand}.

On the other hand, we can also model the dark matter density by assuming that the hot gas is in hydrostatic equilibrium \citep{Chen}. Assuming that the temperature is homogeneous in the small core region of a galaxy cluster and using the single-$\beta$ model, the dark matter density can be expressed by
\begin{equation}
\rho(r)\approx \frac{1}{4\pi r^2}\frac{{\rm d}M(r)}{{\rm d}r}=\frac{3kT_c\beta}{4\pi G\mu m_p}\left[\frac{3r_c^2+r^2}{(r^2+r_c^2)^2}\right],
\end{equation}
where $\mu=0.59$ is the molecular weight and $m_p$ is the proton mass. Both the NFW profile and hydrostatic profile would be considered in our analysis. The relevant parameters for the four target galaxy clusters are listed in Table \ref{Table1}.

The magnetic field profile of a galaxy cluster generally traces the thermal electron density profile. Its strength within such a small cool-core region can be approximately regarded as a constant and is commonly modeled by the following form \citep{Govoni, Kunz}
\begin{equation}
B_0=11\epsilon^{-1/2}\left(\frac{n(r)}{0.1\, {\rm cm^{-3}}}\right)^{1/2}\left(\frac{T_c}{2\, {\rm keV}}\right)^{3/4}\mu {\rm G},
\label{B0}
\end{equation}
where $\epsilon=0.5-1$. In order to obtain the maximum power gain of $L$, we obtain the maximum $B_0$ values of the four galaxy clusters (see Table \ref{Table1}).

Generally speaking, the observed hot gas X-ray luminosity is dominated by thermal bremsstrahlung emission of hot gas particles. The luminosity within the small region with radius $R$ in the cool-core region of a galaxy cluster is given by \citep{Chan1}
\begin{equation}
L_0=\Lambda_0 T_c^{1/2}\int_0^R [n(r)]^2 (4\pi r^2) dr,
\label{L0}
\end{equation}
where $\Lambda_0=1.4\times10^{-27}$ erg cm$^3$ s$^{-1}$ K$^{-1/2}$ and $T_c$ is the core temperature in the unit of K. In general, the X-ray luminosity for a whole galaxy cluster can be as large as $10^{44}$ erg s$^{-1}$. However, since we only consider a very small region in the cool core with $R<3$ kpc for each galaxy cluster, the X-ray luminosity within that small region is just $L_0 \sim 10^{39}-10^{40}$ erg s$^{-1}$ (see Table \ref{Table1}). Since the total power gain in Eq.~(\ref{luminosity}) should not exceed the observed thermal bremsstrahlung luminosity, we can get the upper limits of the PBH fraction $f$ by setting $L \le L_0$.

 In Figs.~1-3, we show the upper limits of $f$ for 3 different mass distributions, mainly for the evaporating PBH mass range. For the monochromatic mass distribution, we can see the conservative limits of $f<1$ for $M_{\rm PBH}$ are less than $\sim 3 \times 10^{16}$ g, which are close to the gamma-ray bounds \citep{Carr,Carr4} and cosmic-ray bounds obtained recently \citep{Boudaud}. However, the constraints obtained in this study are somewhat less stringent than previous constraints (see Fig.~1). Also, $f=1$ is allowed for the non-evaporating PBH mass range $M_{\rm PBH} \ge 10^{17}$ g. For the log-normal and power-law distributions, a large parameter space is ruled out for $f=1$. These constraints are generally more stringent than the previous constraints obtained by using the data of cosmic rays \citep{Boudaud} and gamma rays \citep{Laha2} for the evaporating PBH mass range less than $\sim 10^{17}$ g (see Figs.~2-3 with $\sigma=1$ as the reference for comparison). Overall speaking, the PBH fraction is much less than 1 for a large parameter space for these two extended mass distributions.

\section{Discussion}
 Recent gamma-ray, radio and cosmic-ray data have provided stringent constraints for the PBH fraction in the evaporating PBH mass range \citep{Carr2,Laha,Chan8,Boudaud}. However, not much attention has been paid in using X-ray data to constrain the PBH fraction. In this study, we have examined the X-ray data of some cool-core galaxy clusters to constrain the PBH fraction.  Our result is similar to the previous results that PBHs only constitute a minor component of dark matter for a large parameter space in the evaporating PBH mass range. Although the constraints in our study are less stringent than previous constraints for the monochromatic mass distribution, we get more stringent constraints for the power-law and log-normal PBH mass distribution. However, note that these extended mass distributions are somewhat model-dependent, though they are usually regarded as the benchmark models for consideration. Nevertheless, adding the analysis using X-ray data of galaxy clusters, we can generate a multi-wavelength picture of the PBH fraction constraints, which includes radio waves, X-ray and gamma rays. It can also initiate some new studies using X-ray data of other cool-core galaxy clusters to constrain PBH dark matter.

Generally speaking, our proposed method is an indirect way to constrain the PBH fraction. It relies on the physics of the Coulomb interaction between the thermal hot gas particles and the high-energy particles emitted by the PBHs. Nevertheless, the constraints obtained by this method did not consider other possible heating sources, such as the heating of Active Galactic Nuclei (AGN) and the cooling flow process. These processes could input extra energy to the central regions of the galaxy clusters so that the power gain $L$ in Eq.~(\ref{luminosity}) might be underestimated. As a result, the PBH fraction obtained in this study is a conservative bound.

On the other hand, using upper limits of electromagnetic wave fluxes in different bands (i.e. the multi-wavelength approach) to constrain PBH fraction is mainly based on the Hawking radiation emitted by PBHs.  This method has some limitation in the non-evaporating PBH mass regime ($M_{\rm PBH} \ge 10^{17}$ g). The temperature of a black hole becomes low for large PBH mass so that the energy of the particles emitted would quickly cool down to non-relativistic. These particles would eventually be degenerated with the thermal particles in the interstellar medium. Hence, the limits of PBH fraction would be difficult to obtain. Unfortunately, our method also suffers from the same problem and it can be applied for constraining the evaporating PBH mass range only. For the non-evaporating PBH mass regime, there are other methods such as lensing which can give some limited constraints for that regime \citep{Smyth,Carr4}. More new methods have to be developed to explore this regime.

\begin{figure}
\vskip 3mm
\includegraphics[width=120mm,angle=-90]{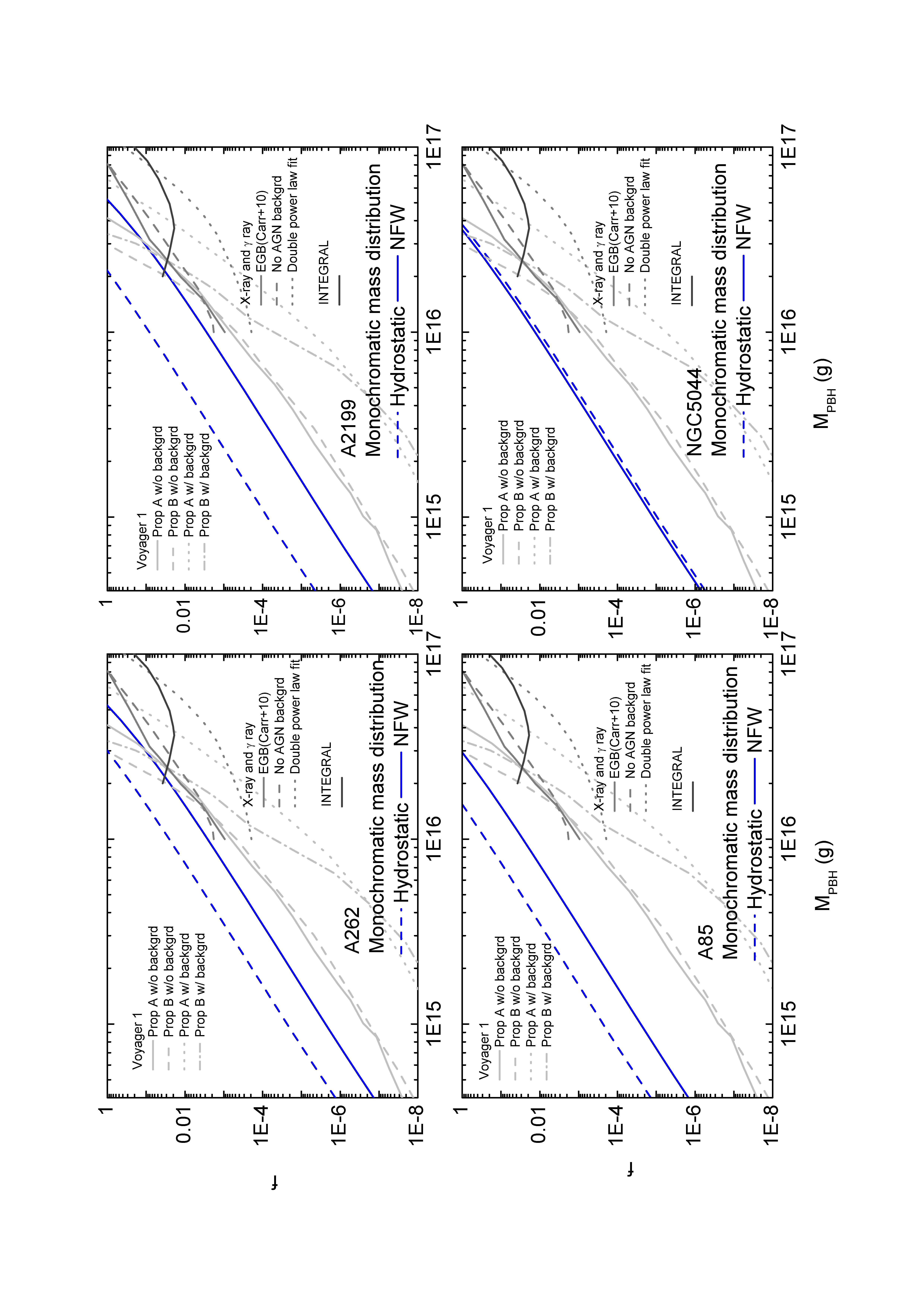}
\caption{The upper limits of the PBH fraction $f$ for the four cool-core clusters (A262, A2199, A85, and NGC5044) as a function of the PBH monochromatic mass. The gray lines indicate the upper limits of $f$ constrained by the cosmic-ray, gamma-ray, and X-ray data \citep{Ballesteros,Boudaud,Laha}.}
\label{Fig1}
\end{figure}

\begin{figure}
\vskip 3mm
\includegraphics[width=120mm,angle=-90]{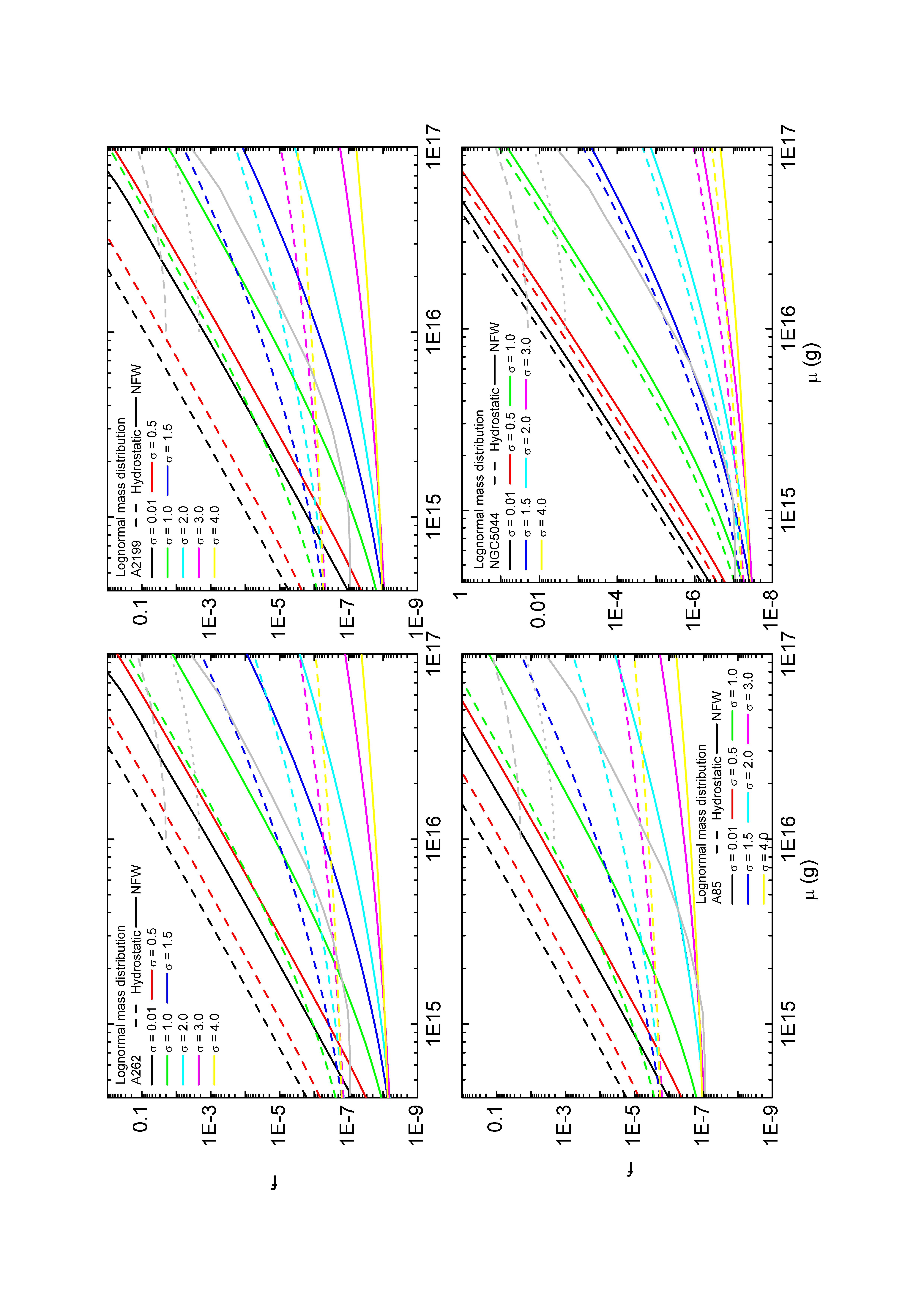}
\caption{The colored lines represent the upper limits of $f$ for the four cool-core clusters as a function of the median PBH mass $\mu$ for the log-normal mass distribution. The gray solid lines indicate the previous constraints from the Voyager I observations with $\sigma=1$ \citep{Boudaud}. The previous constraints from the Milky Way gamma-ray data in \citet{Laha2} are shown by the gray dashed lines (isothermal density profile with $\sigma=1$) and the gray dotted lines (NFW density profile with $\sigma=1$).}
\label{Fig2}
\end{figure}

\begin{table}
\caption{The temperature of the cool-core region $T_c$ within a small radius $R$ \citep{Sanderson} and the other parameters \citep{Chen, Chan1} used in our calculation
for the four cool-core galaxy clusters. The central magnetic field $B_0$ and the luminosity $L_0$ within the cool-core regions are obtained by Eq.(\ref{B0}) and Eq.(\ref{L0}), respectively.}

\begin{tabular}{ |l|l|l|l|l| }
 \hline\hline
    &A262    &A2199   &A85  &NGC5044 \\
 \hline
 $T_c$(keV)  & $<1$  &   $<2$        &$<3$       &$<0.8$    \\
 $\rho_s(10^{14}M_{\odot}$Mpc$^{-3})$&14.1&9.56&8.34&14.7\\
 $r_s$(kpc)&172&334&444&127\\
 $R$ (kpc)&2&3&3&1 \\
 $z$ & 0.0161&0.0302 & 0.0556& 0.0090 \\
 $\beta$   &   $0.433_{-0.017}^{+0.018}$     &  $0.665_{-0.021}^{+0.019}$       &  $0.532_{-0.004}^{+0.004}$    & $0.524_{-0.003}^{+0.002}$         \\
 $r_c$(kpc)   &  $30_{-7}^{+8}$      &$102_{-7}^{+7}$        &$60_{-2}^{+2}$     &$8_{-0}^{+0}$         \\
 $n_0$($10^{-2}$ cm$^{-3}$)    & $0.94_{-0.10}^{+0.15}$       & $0.97_{-0.03}^{+0.03}$       &$3.00_{-0.12}^{+0.12}$     &$4.02_{-0.03}^{+0.03}$         \\
$B_0(\mu {\rm G})$ &2.9&4.9&11.6&5.0\\
$L_0$(erg s$^{-1}$)&$<5.6\times10^{38}$&$<2.3\times10^{39}$ &$<2.7\times10^{40}$& $<6.3\times10^{38}$\\
  \hline\hline
\end{tabular}
\label{Table1}
\end{table}

\section{acknowledgements}
The work described in this paper was supported by a grant from the Research Grants Council of the Hong Kong Special Administrative Region, China (Project No. EdUHK 28300518).

\end{document}